\documentclass[10pt]{llncs}
\usepackage{llncsdoc}
\usepackage{epsfig}
\begin{document}
\pagestyle{headings}
\mainmatter

\title{\bf Self-Sustained Thought Processes
in a Dense Associative Network}

\titlerunning{Self-Sustained Thought Processes}

\author{Claudius Gros }

\institute{Institut f\"ur Theoretische Physik,
Universit\"at Frankfurt,
Max-von-Laue-Strasse 1,
60438 Frankfurt am Main, Germany
\footnote[1]{Address at time of submission:
Institut f\"ur Theoretische Physik,
Universit\"at des Saarlandes,
66041 Saarbr\"ucken, Germany.}
          }

\date{March 16, 2005}
\maketitle

\begin{abstract}
% \normalsize
Several guiding principles for thought processes 
are proposed and a neural-network-type
model implementing these principles is presented
and studied. We suggest to consider thinking
within an associative network built-up of
overlapping memory states. We consider a
homogeneous associative network as biological
considerations rule out distinct conjunction
units between the information (the memories)
stored in the brain.
We therefore propose that memory states have
a dual functionality: They represent on one side
the stored information and serve, on the other
side, as the associative links in between the different
dynamical states of the network which consists
of transient attractors.

We implement these principles within a generalized
winners-take-all neural network with sparse coding
and an additional coupling to local reservoirs.
We show that this network is capable to generate
autonomously a self-sustained time-series of memory states
which we identify with a thought process. Each memory state
is associatively connected with its predecessor.

This system
shows several emerging features, it is
able (a) to recognize external patterns
in a noisy background, (b) to focus attention
autonomously and (c) to represent hierarchical memory
states with an internal structure.
\end{abstract}

%%%%%%%%%%%%%%%%%%%%%%%%%%%
%%%%%%%%%%%%%%%%%%%%%%%%%%%

% \newpage
% \tableofcontents
% \newpage

%%%%%%%%%%%%%%%%%%%%%%%%%%%
%%%%%%%%%%%%%%%%%%%%%%%%%%%

\section{Introduction}

The notion of `thinking' comes in various
flavors. We may associate thinking with
logical reasoning or with a series of
associative processes. The latter activity
is performed effortless by the human brain
and is at the center of the investigation we will
present here. We will consider in particular
associative processes which may occur also
in the absence of any interaction of the brain
with the outside world. It is clear that without
any prior stored information - our memories -
this kind of thought process would be semantically empty.
We do therefore investigate the autonomous
generation of a time-series of memory-states
by a cognitive system, being it natural or
artificial.

We consider consequently thought processes to be
characterized by the spontaneous activation
of one memory state by another one, leading to
a history of memory states. This process should
be autonomous and no outside regulative unit 
should be needed in order to control this 
dynamical process. In order to make sense,
each activated memory state should be closely 
associated to its predecessor. A history of
random memory states would not possibly classify as
a true thought process.

A key question in this context is then: When 
can two or more memory states be considered to
be closely associated? Intuitively this is not
a problem: Remembering a trip to the forest with
our family for a picnic we may associate this activity with a 
trip to the forest to cut a Christmas tree. These
two memory-states are intuitively related. 
When we store new memory states, like the two trips to
the forest in the above example, in our brain,
the appropriate associative links need to be generated
spontaneously. But how should our brain be capable
of finding all possible relations linking this new information
associatively with all previously stored memory-states? 
An exhaustive search would not be feasible, the time 
needed to perform it would be immense.

This computational problem of embedding new memory states 
into their relevant semantic context would not occur if no
explicit associative links would be needed at all.
This can be achieved when considering
networks with overlapping memory states.
In this case no additional conjunction units describing
associative links in between two stored memories
are needed. These associative links would be
formed by other memories. Any new information
learned by the network then acquires naturally
meaningful associative links whenever it shares
part of its constituent information with other
memory states.

We do therefore consider a homogeneous network,
with only one kind of constituent building block:
the memories themselves.
The memory states then show a dual functionality: 
depending on the initial condition, an associative link in between
two or more activity centers could be either a stationary
memory state by itself or it could serve to form an 
association in between two sequential memory states in the 
course of a thought process.

We propose a generalized neural-network model 
capable of simulating the here defined kind of 
thought processes. We do not claim
that actual thought processes in biological cybernetic systems
(in our brain for instance) will be described accurately
by this model. However, the kind of thought processes 
proposed here seem to be a mandatory requirement if
a homogeneous associative network without an external
regulative unit wants to acquire true
information-processing capabilities.
In such kind of networks the information-processing must
be self-organized by an autonomous dynamical process. 
From a functional point of view it is evident that this
self-regulated information processing needs 
to be implemented in biological 
cognitive systems, like the human brain, in one
way or another.

We note that these self-organized association processes work
only within dense associative networks, where essentially all
activity centers are connected among themselves, 
forming what one calls in network-theory a
`Giant Strongly Connected-Component'
% (Dorogovtsev \& Mendes, 
\cite{dorogovstsev03}.
In a sparse network there would be 
many unconnected subclusters incapable to 
communicate autonomously. Therefore we consider here
dense and homogeneous associative 
networks (dHAN) and one might argue that 
the human brain does fall into this category.

%%%%%%%%%%%%%%%%%%%%%%%%%%%%%%%%%%
 \begin{figure}[t]
 \centerline{
 \epsfig{file=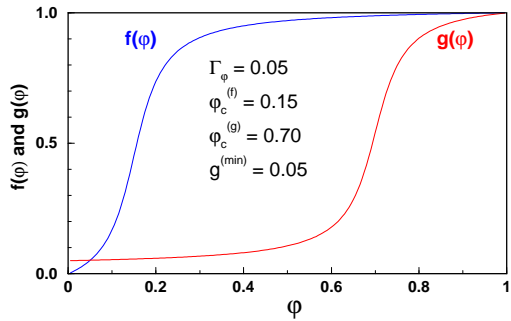,width=0.55\textwidth,angle=0}
            }
 \caption{\small Illustration of the
 reservoir-functions $f(\varphi)$ and $g(\varphi)$
 as defined by Eq.\ (\ref{eq_fg})
 for $ \varphi_c^{(f)}=0.15$, $ \varphi_c^{(g)}=0.7$,
 $\Gamma_\varphi=0.05 $ and
 $ g^{(min)}=0.05$
 \label{fig_fg}
         }
 \end{figure}
%%%%%%%%%%%%%%%%%%%%%%%%%%%%%%%%%%

%%%%%%%%%%%%%%%%%%%%%%%%%%%%%%%%%%
\begin{table}[b]
\caption{\small Sets of model-parameters used for the simulations
presented here. 
$w$/$z$ denote the non-zero matrix elements of the link-matrices
$w_{i,j}$/$z_{i,j}$ entering Eq.\ (\ref{xdot}). The filling/depletion
rates for the reservoir $ \Gamma_\varphi^\pm$ and $x_c$ enter
Eq.\ (\ref{phidot}). The critical reservoir-levels for inhibition
and activation, $ \varphi_c^{(f/g)}$ enter Eq.\ (\ref{eq_fg}),
as well as the width $\Gamma_{\varphi}$ for the reservoir function
and the minimal values $f^{(min)}=0$ 
and $g^{(min)}$
       }
\begin{center}
\begin{tabular}{cccccccccc}
\hline
$\qquad $ & $\quad w\quad $ & $\quad z\quad $ & $ \quad x_c \quad $ &
$\quad \Gamma_\varphi^+\quad $ 
   & $\quad \Gamma_\varphi^-\quad $ 
	& $\quad \varphi_c^{(f)}\quad $ 
	& $\quad \varphi_c^{(g)}\quad $
	& $\quad \Gamma_{\varphi}\quad $
	& $\quad g^{(min)}\quad $ \\
\hline
(a) & 0.15 & -1.0 &  0.85& 0.004 & 0.009 & 0.15 & 0.7 & 0.05 & 0.00\\
(b) & 0.15 & -1.0 &  0.50& 0.005 & 0.020 & 0.15 & 0.7 & 1.00 & 0.10\\
\hline
\end{tabular}
\end{center}
  \label{tab_parameters}
\end{table}
%%%%%%%%%%%%%%%%%%%%%%%%%%%%%%%%%%

%%%%%%%%%%%%%%%%%%%%%%%%%%%
%%%%%%%%%%%%%%%%%%%%%%%%%%%

\section{Associative and Homogeneous Networks}
\label{sect_model_short_term}

We consider an associative network with 
$N$ sites, which we also call activity centers (AC). 
Each AC represents some specific biologically relevant
information, normally in a highly preprocessed
form 
%(see Hubel \& Wiesel, 
\cite{hubel65}.
Examples are ACs for colors, shapes, distances,
movements, sounds and so on.
Each AC is characterized by the individual 
activity $x_i\in[0,1]$ ($i=1,\dots, N$).
An AC is active when $x_i$ is close to one. 
We identify ensembles of active ACs with memory 
states 
% (see McLeod, Plunkett \& Rolls, 
\cite{mcLead98}.

In addition to the activity levels $x_i$ we introduce
for every AC a new variable $\varphi_i\in[0,1]$, 
which characterizes the level of the individual
activity reservoirs. This variable plays, as it will 
become clear from the discussions further below,
a key role in facilitating the self-sustained thought
process and distinguishes the dHAN from standard
neural networks
%(M\"uller, Reinhardt \& Strickland, 
\cite{mue95}.

We consider here a continuous-time ($t$) evolution,
$x_i=x_i(t)$ and $\varphi_i=\varphi_i(t)$. The
differential equations ($\dot x_i(t)={d\over dt}x_i$) 

\begin{eqnarray} \label{xdot}
\dot x_i &=& (1-x_i)\Theta(r_i)r_i \,+\, x_i[1-\Theta(r_i)]r_i \\
     r_i &=& b_i\, + \,
	        g(\varphi_i)\sum_{j=1}^N w_{i,j}x_j
	       \,+\,\sum_{j=1}^N z_{i,j}x_jf(\varphi_j)
\label{ri} \\ \label{phidot}
\dot\varphi_i & =&  \Gamma_\varphi^+\,\Theta(x_c-x_i)(1-\,\varphi_i)
\,-\, \Gamma_\varphi^-\,\Theta(x_i-x_c)\,\varphi_i~. 
\end{eqnarray}
determine the time-evolution of all $x_i(t)$.
Here the $r_i$ are growth rates 
and the $b_i$ the respective biases.\footnote{The differential equations
(\ref{xdot}) and (\ref{ri})
are akin to the Lotka-Volterra equations
discussed by Fukai and Tanaka \cite{fukai97}.
         }
We will discuss the role of the bias further below, 
for the time being we consider
$b_i\equiv0$, if not stated otherwise.\footnote{The time-unit 
is arbitrary in principle and could be tuned,
as well as most of the parameters entering 
Eqs.\ (\ref{xdot}) and\ (\ref{ri}), in order to
reproduce neurobiologically observed time-scales.
For convenience one could take a millisecond for the
time unit, or less.
         }

The function $\Theta(r)$ occurring in Eq.\ (\ref{xdot}) is the
step function: $\Theta(r)=1,0$ for $r>0$ and $r<0$ respectively.
The dynamics, Eqs.\ (\ref{xdot}) and (\ref{phidot}),
respects the normalization $x_i\in[0,1]$ and 
$\varphi_i\in[0,1]$
due to the prefactors $(1-x_i)$, $(1-\varphi_i)$
and $x_i$, $\varphi_i$  
for the growth and depletion processes.

The neural-network-type interactions
in between the activity centers are given by
the matrices $0\le w_{i,j}\le w$ and $z_{i,j}\le -|z|$ for
excitatory and inhibitory connections respectively. 
The breakdown of the link-matrix in an excitatory and 
inhibitory sector can be considered as a reflection 
of the biological observation that excitatory and 
inhibitory signals are due to neurons and interneurons 
respectively. Any given connection is
either excitatory or inhibitory, but not both at the same time:
$w_{i,j}z_{i,j}\equiv0$, for all pairs ($i,j$). 
We do not consider here self-interactions
(auto-associations): $w_{i,i}=z_{i,i}\equiv0$.

%%%%%%%%%%%%%%%%%%%%%%%%%%%%%%%%%%
\begin{figure}[t]
\centerline{
\epsfig{file=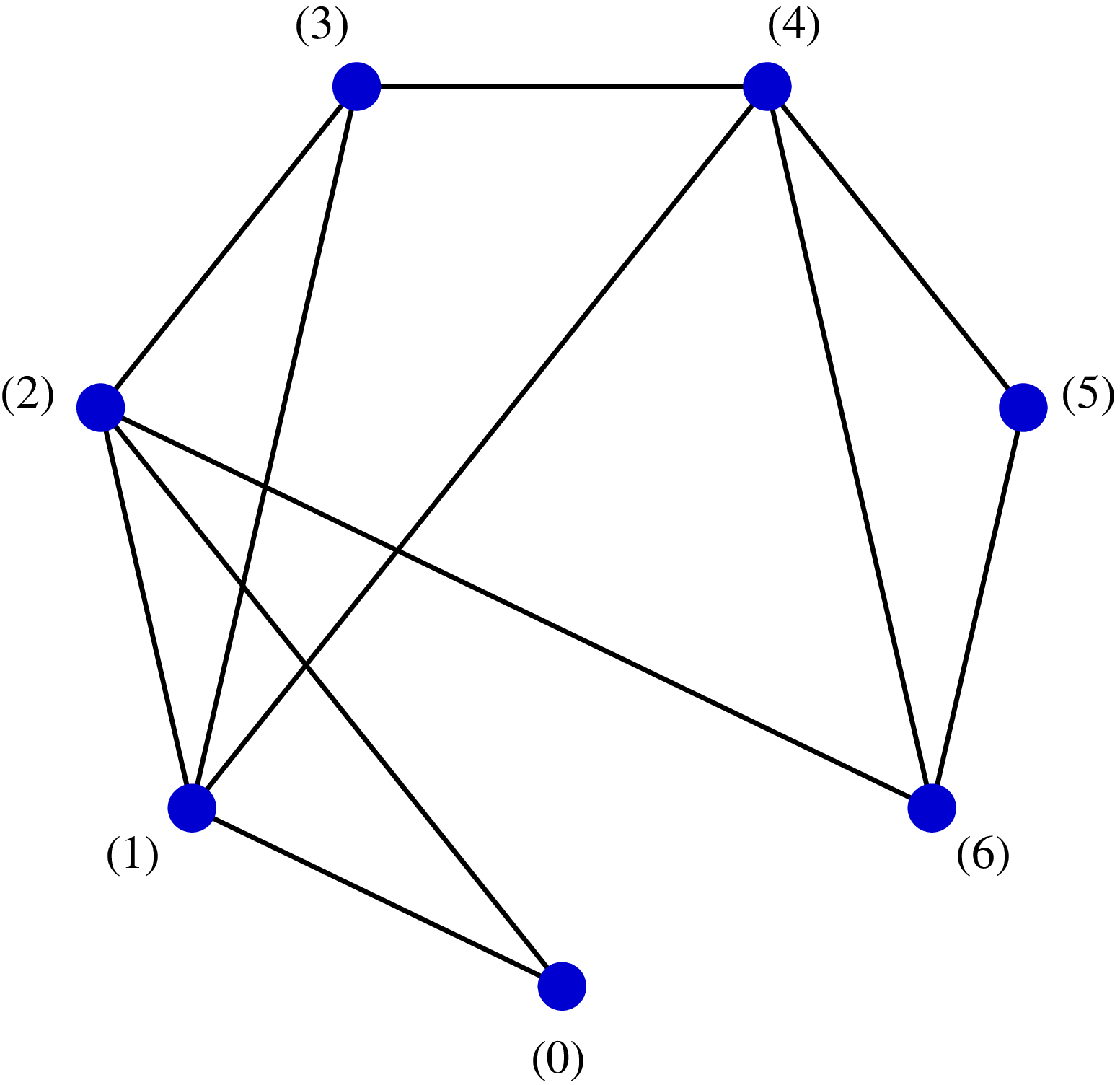,width=0.35\textwidth,angle=0}\hspace{5ex}
\epsfig{file=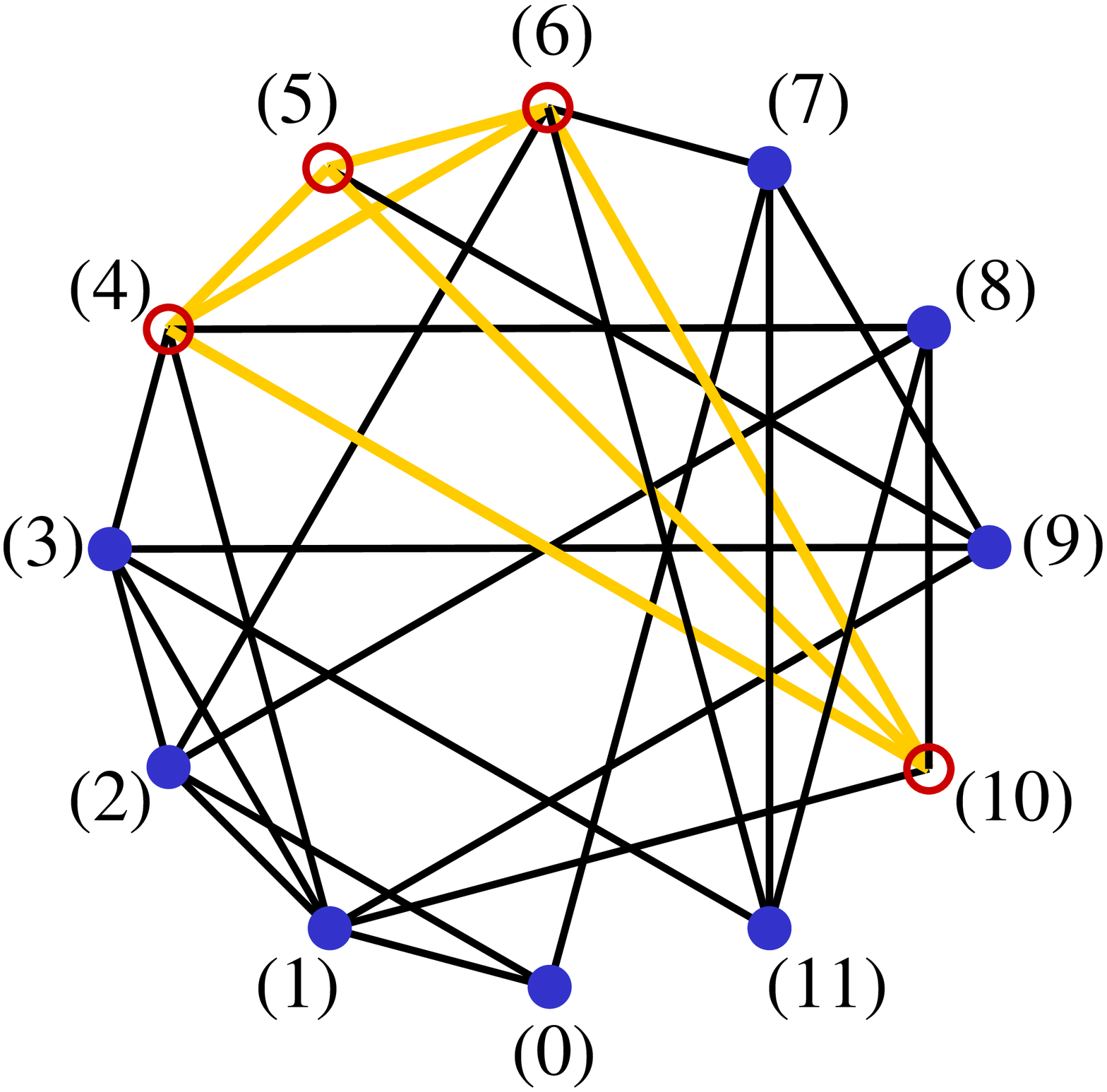,width=0.35\textwidth,angle=0}
           }
\caption{\small Two small networks for illustrational purposes. 
The arrangement of the activity centers (filled blue circles) 
is arbitrary, here we have chosen a circular 
arrangement for a good overview.
Shown are the excitatory links (the non-zero matrix elements of
$w_{i,j}$, black lines). Two activity centers 
not connected by $w_{i,j}$ are inhibitorily 
connected via $z_{i,j}$.
\newline
Left: A seven-center network with five stable memory states:
(0,1,2), (1,2,3), (1,3,4), (4,5,6) and (2,6). \newline
Right: A 12-center network with 7 2-center memory states,
7 3-center memory states and one 4-center memory state
(which is highlighted). It contains a total of 28 links
(non-zero matrix-elements of $w_{i,j}$)
		  \label{fig_illustration}
        }
\end{figure}
%%%%%%%%%%%%%%%%%%%%%%%%%%%%%%%%%%

We consider here the recurrent case with
$w_{i,j} = w_{j,i}$ and $z_{i,j} = z_{j,i}$, but the
model works fine also when this symmetry is partially
broken. This will happen anyhow dynamically via the
reservoir-functions $f(\varphi)$ and $g(\varphi)$.
These functions govern the
interaction in between the activity levels $x_i$ and
the reservoir levels $\varphi_i$. 
They may be chosen as washed-out step functions
of a sigmoidal form like

\begin{equation} \label{eq_fg} %\textstyle
g(\varphi)\ =\ g^{(min)} \,+\, 
\left(1.0-g^{(min)}\right)
{
{\rm atan}[(\varphi-\varphi_c^{(g)})/\Gamma_\varphi] -
{\rm atan}[(0-\varphi_c^{(g)})/\Gamma_\varphi]
\over 
{\rm atan}[(1-\varphi_c^{(g)})/\Gamma_\varphi] -
{\rm atan}[(0-\varphi_c^{(g)})/\Gamma_\varphi]
}~,
\end{equation}
with a suitable width $\Gamma_{\varphi}$.
For an illustration see Fig.\ \ref{fig_fg}.
The effect of the reservoir functions depends
on the value of the respective reservoir-levels
$\varphi_i$, which are governed by Eq.\ (\ref{phidot}).

For $x_i>x_c$ (high activity level) the reservoir-level
$\varphi_i$ decreases with the rate $\Gamma_\varphi^-$. For
$x_i<x_c$ (low activity level) the reservoir-level
$\varphi_i$ increases with the rate $\Gamma_\varphi^+$. 
A low reservoir level will have two effects: The ability to
suppress another activity center via an
inhibitory link $z_{i,j}$, which will be reduced by 
$f(\varphi_i)\in[0,1]$ and the activation by other centers 
via an excitatory link $w_{i,j}$, which will
be reduced by $g(\varphi_i)\in[0,1]$, see Eq.\ (\ref{ri}). 

\begin{figure}[t]
\centerline{
\epsfig{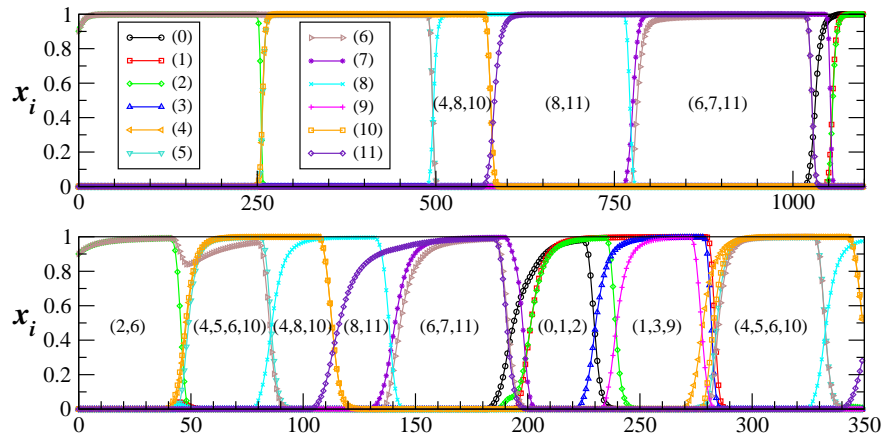} 
           }
\smallskip
\caption{\small The activity $x_i(t)$ 
for the thought process
$(2,6)\rightarrow(4,5,6,10)
\rightarrow (4,8,10)\rightarrow(8,11)
\rightarrow (6,7,11)\rightarrow(0,1,2)
\rightarrow (1,3,9)\rightarrow(4,5,6,10)
$ for the 12-site network in Fig.\ \ref{fig_illustration}.
Top/Bottom: Using the parameter sets (a)/(b) of
Table\ \ref{tab_parameters}. Note the different scaling
for the respective time-axis
		  \label{fig_12_AA}
        }
\end{figure}

The dynamics induced by Eq.\ (\ref{xdot}) leads to a relaxation
towards the next stable memory state within a short time-scale
of 
$\Gamma_r^{-1}\approx\left|w_{i,j}\right|^{-1}\approx
\left|z_{i,j}\right|^{-1}$ 
(for the non-zero matrix-elements of the 
link-matrices). Choosing the rates $\Gamma_\varphi^\pm$
for the reservoir dynamics to be substantially smaller than
the relaxation-rates  $\Gamma_r$ we obtain a separation of
time-scales for the stabilization of memory states and for the
depletion/filling of the activity reservoirs $\varphi_i(t)$
described by Eq.\ (\ref{phidot}).
This separation of time-scales is evident in
the simulations presented in Fig.\ \ref{fig_12_AA}.
For illustrational purposes we present the activity-levels
for a (very) small system, the 12-center network 
illustrated in Fig.\ \ref{fig_illustration}. 
The time-scales of the dynamics are however system-size
independent. We note, that only a finite
number of centers are active at any given time.
Before we  discuss the dynamics of the thought process
more in detail in Sect.\ \ref{Dynamical_thought_processes},
we will take a closer look at the nature of the
transient attractor stabilized for short time-scales.

\subsection{Memory States}
\label{Memory_states}

We consider here memory states which contain
only a finite number, typically between 2 and 7, 
of constituent activity centers. 
This is a key difference between the dHAN
investigated here and standard neural networks, 
where a finite fraction of all neurons might
be active simultaneously 
%(O'Reilly, 
\cite{oReilly98}.

The stabilization of memory states made up of clusters
with a finite number $Z=2,3,\dots$ of
activity centers is achieved by an
inhibitory background of links:

\begin{equation}
z_{i,j}\le z < 0, \qquad\qquad
\forall\,(w_{i,j} =0, i\ne j)~.
\label{eq_take_all}
\end{equation}
In Fig.\ \ref{fig_illustration} we illustrate a
7-center network. Illustrated in Fig.\ \ref{fig_illustration}
by the black lines are the excitatory links, i.e.\ the non-zero
matrix-elements of $w_{i,j}$. All pairs $(i,j)$ of 
activity-centers not connected
by a line in Fig.\ \ref{fig_illustration} have
$z_{i,j}\le -|z|$. If $|z|$ is big enough, then only
those clusters of activity centers are dynamically
stable, in which all participating centers are mutually
connected. 

To see why, we consider an AC (i) outside a
Z-center memory state (MS). The site (i)
cannot have links (finite $w_{i,j}$)
to all of the activity centers (j)
making up this memory state. Otherwise (i) would
be part of this MS. There are
therefore maximally $Z-1$ positive
connections in between (i) and the MS.
The dynamical stability of the memory state 
is guaranteed if the total link-strength
between (i) and the MS is not too strong:
\begin{equation}
|z|\ >\ \sum_{j\in {\rm MS}} w_{i,j}, \qquad \quad
|z|\ >\ (Z-1)\,w~,
\label{eq_Z}
\end{equation}
where the second equation holds for the
uniform case, $w_{i,j}\equiv w>0$.

For an illustration of this relation we consider 
the case $x_3=x_4=x_5=x_6=0$
and $x_0=x_1=x_2=1$ for the 7-center network
of Fig.\ \ref{fig_illustration}. The growth-rate
for center (3) is then: $r_3 = 2w-|z|$. For
$2w-|z|>0$ center (3) would start to become active
and a spurious state (1,2,3,4) would result. Taking
$2w-|z|<0$ both (0,1,2) and (1,2,3) are stable
and viable 3-center memory states. 

A `spurious memory state' of the network
would occur if a group of ACs remains
active for a prolonged time
even though this grouping does not
correspond to any stored memory state.
No such spurious memory state is dynamically
stable when Eq.\ (\ref{eq_Z}) is fulfilled.
For the simulation
presented here we have chosen $|z|/w>6$. This
implies that
memory states with up to $Z=7$ activity centers 
are stable, see Table \ref{tab_parameters}
and Eq.\ (\ref{eq_Z}). 

This kind of encoding of the link-matrices is called
a `winners-take-all' situation\footnote{Our 
winners-take-all setting differs from the
so-called `K-winners-take-all' configuration 
in which the $K$ most active neurons suppress 
the activities of all other neurons via an inhibitory
background 
%(see Kwon \& Zervaksi, 
\cite{kwon95}.
         }, since fully
interconnected clusters will stimulate each other 
via positive intra-cluster $w_{i,j}$. There will be
at least one negative $z_{i,j}$-link in between
an active center of the winning memory state to
every out-of-cluster AC, suppressing in this way the 
competing activity of all out-of-cluster
activity centers. 

%%%%%%%%%%%%%%%%%%%%%%%%%%%%%%%%%%
%%%%%%%%%%%%%%%%%%%%%%%%%%%%%%%%%%

\subsection{Hierarchical Memory States}
\label{sect_hierarchical_memory_states}

In the above discussion we have considered in part
the uniform case $w_{i,j}\equiv w$ for all 
non-zero excitatory links. In this case,
all features making up a memory state are bound
together with the same strength. Such a memory
state has no internal structure, it is just
the reunion of a bunch of semantic nodes with
no additional relations in between them.
Memory states
corresponding to biological relevant objects
will however exhibit in general a hierarchical structure
%(Riesenhuber \& Poggio, 
\cite{riesenhuber99} 
% Mel \& Fiser, 
\cite{mel00}. Let us give an
example: A memory state denoting a `boy' 
may involve a grouping of ACs 
corresponding to (face), (shirt), (pants), (legs),
(red), (green) and so on. This memory state
is well defined in our model whenever there are
positive links $w_{i,j}>0$ in between all of
them. 

There is now the need for additional information 
like: is `red' the color of the shirt or of the trousers?
That is, there is the need to `bind' the color
red preferentially to one of the two pieces of
clothes.
It is possible to encode this internal information
into the memory state `boy' 
(face,shirt,pants,legs,red,green,..)
by appropriate modulation of the internal connections.
In order to encode for instance that
red is the color of the shirt, one sets the link
(red)-(shirt) to be much stronger than the
link (red)-(pants) or (red)-(legs). This is perfectly
possible and in this way the binding of
(red) to (shirt) is achieved. The structure of the 
memory states defined here for the dHAN is therefore 
flexible enough to allow for a (internal) 
hierarchical object representation.

No confusion regarding the colors of the shirt
and of the pants arises in the above example
when variable link-strengths $w_{i,j}$ are used. 
Note however, that this is possible
only because small and negative links $z_{i,j}$
are not allowed in our model, a key difference to
the most commonly used neural-network models.
If weak inhibitory links would
be present, the boundary of memory states could
not be defined precisely. There would be no
qualitative difference in between a small negative
and a small positive synapsing strength. Furthermore,
the stability condition Eq.\ (\ref{eq_Z}) would break down.

%%%%%%%%%%%%%%%%%%%%%%%%%%%%%%%%%%
\begin{figure}[t]
\centerline{
\epsfig{file=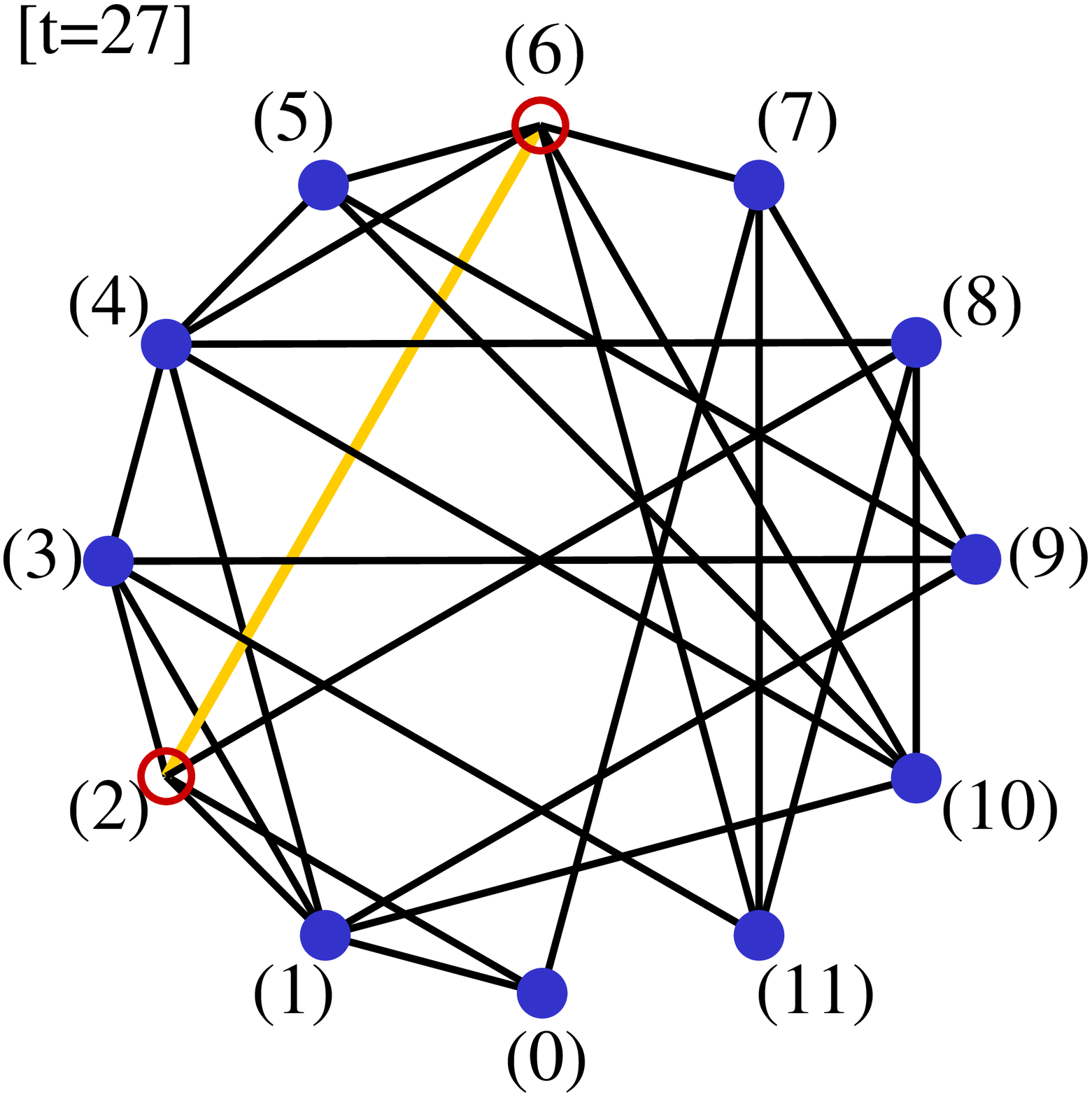,width=0.25\textwidth,angle=0} \hspace{1ex}
\epsfig{file=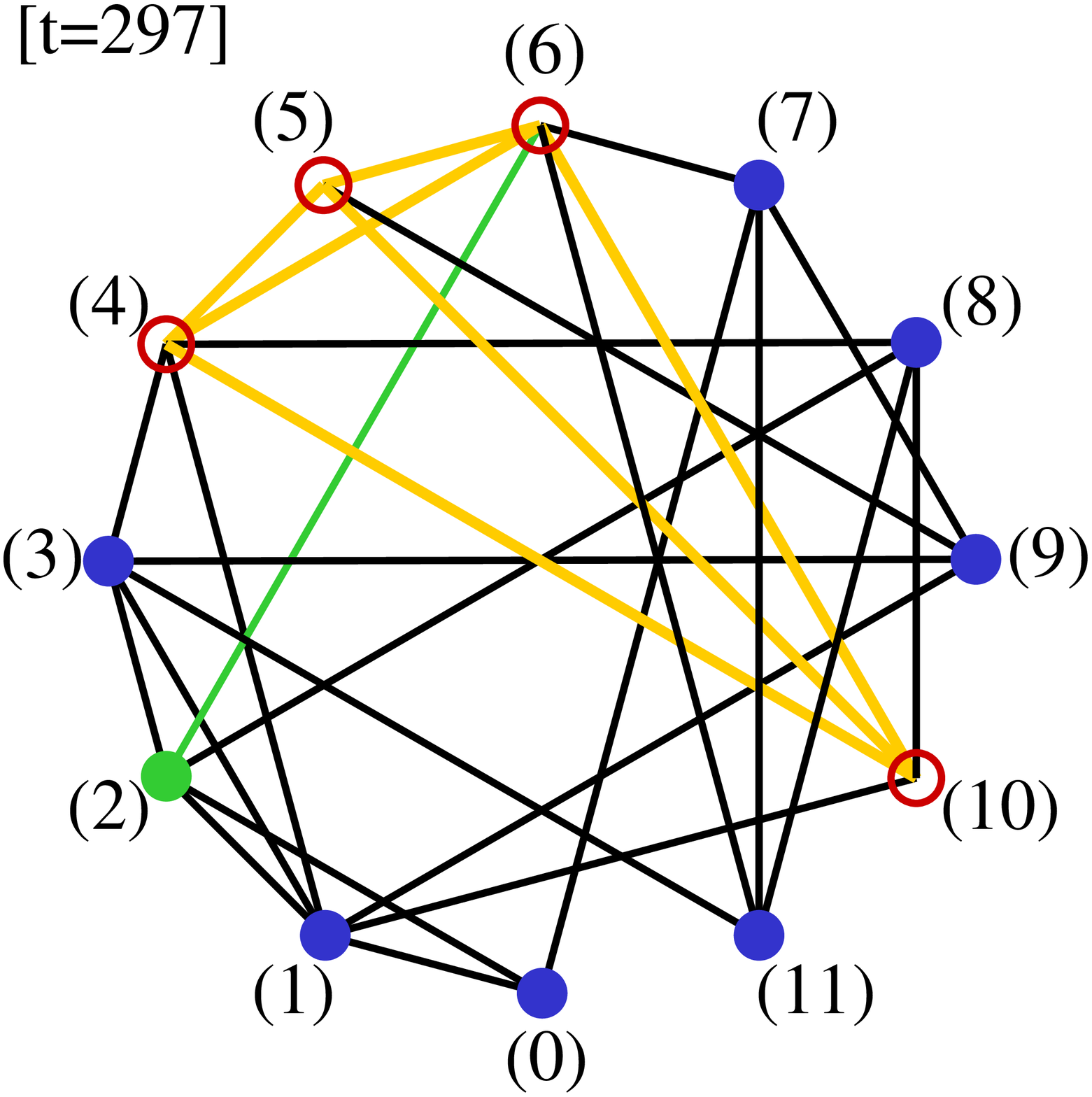,width=0.25\textwidth,angle=0} \hspace{1ex}
\epsfig{file=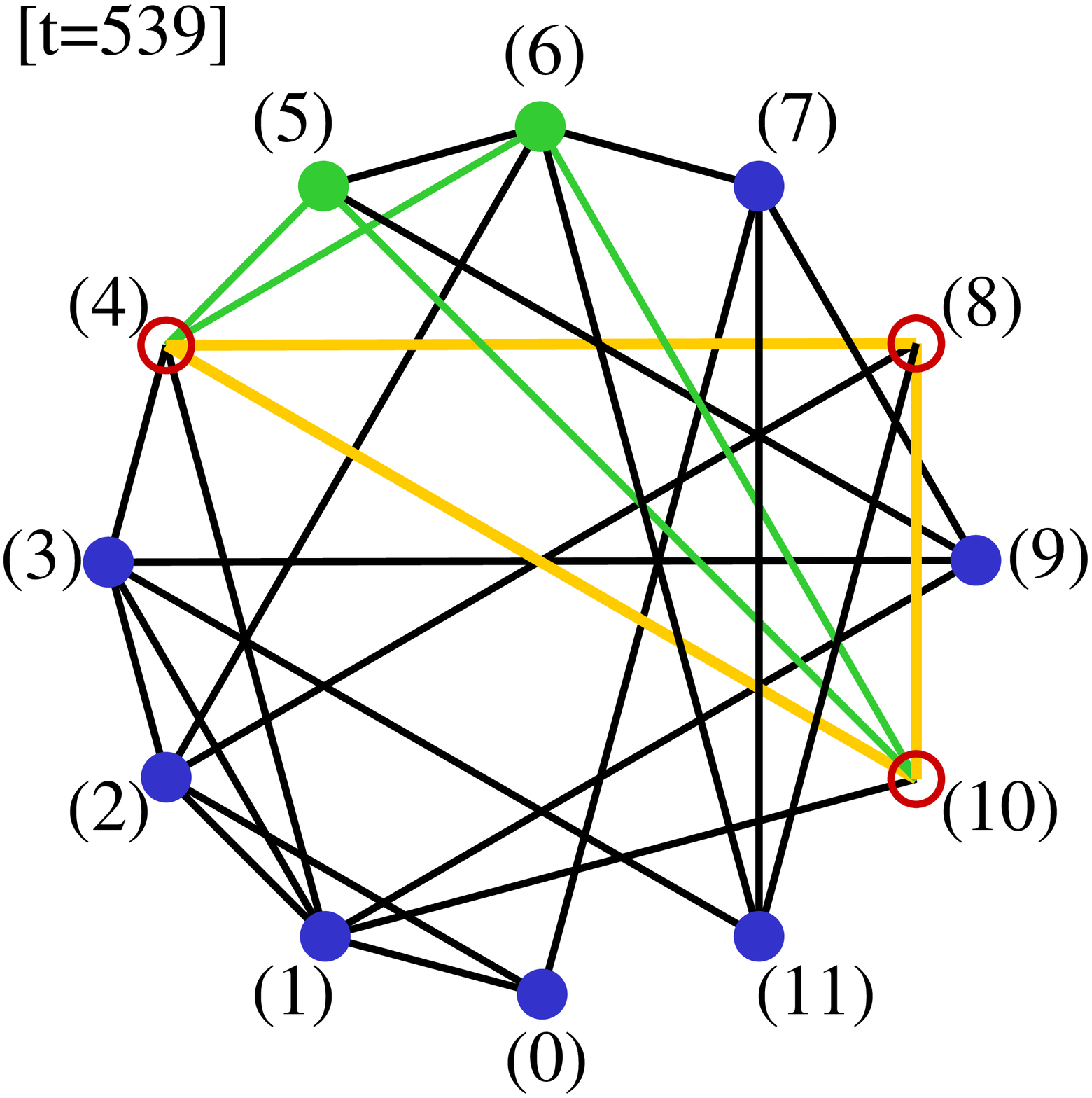,width=0.25\textwidth,angle=0} 
           }
\vspace{2ex}

\centerline{
\epsfig{file=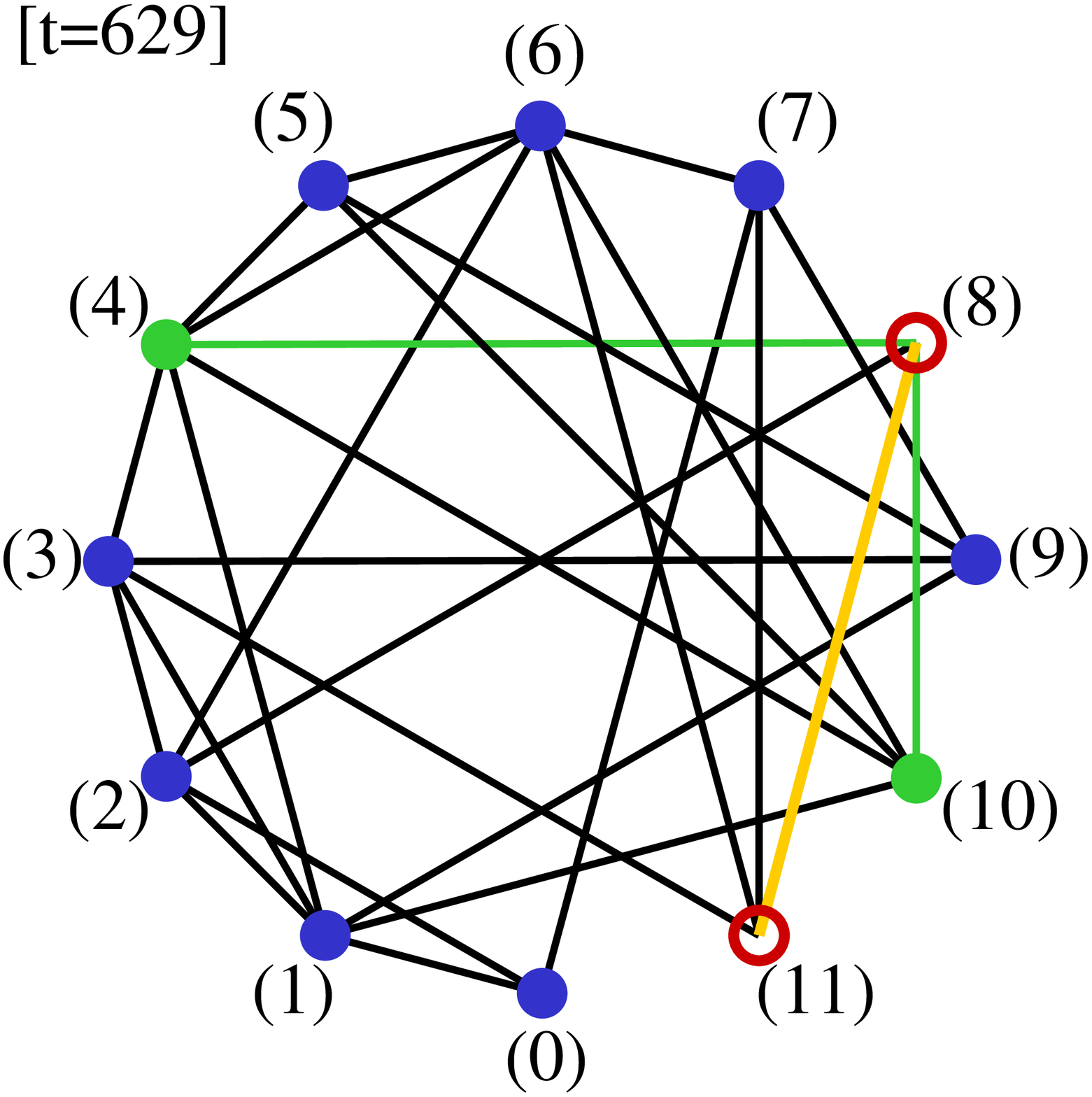,width=0.25\textwidth,angle=0} \hspace{1ex}
\epsfig{file=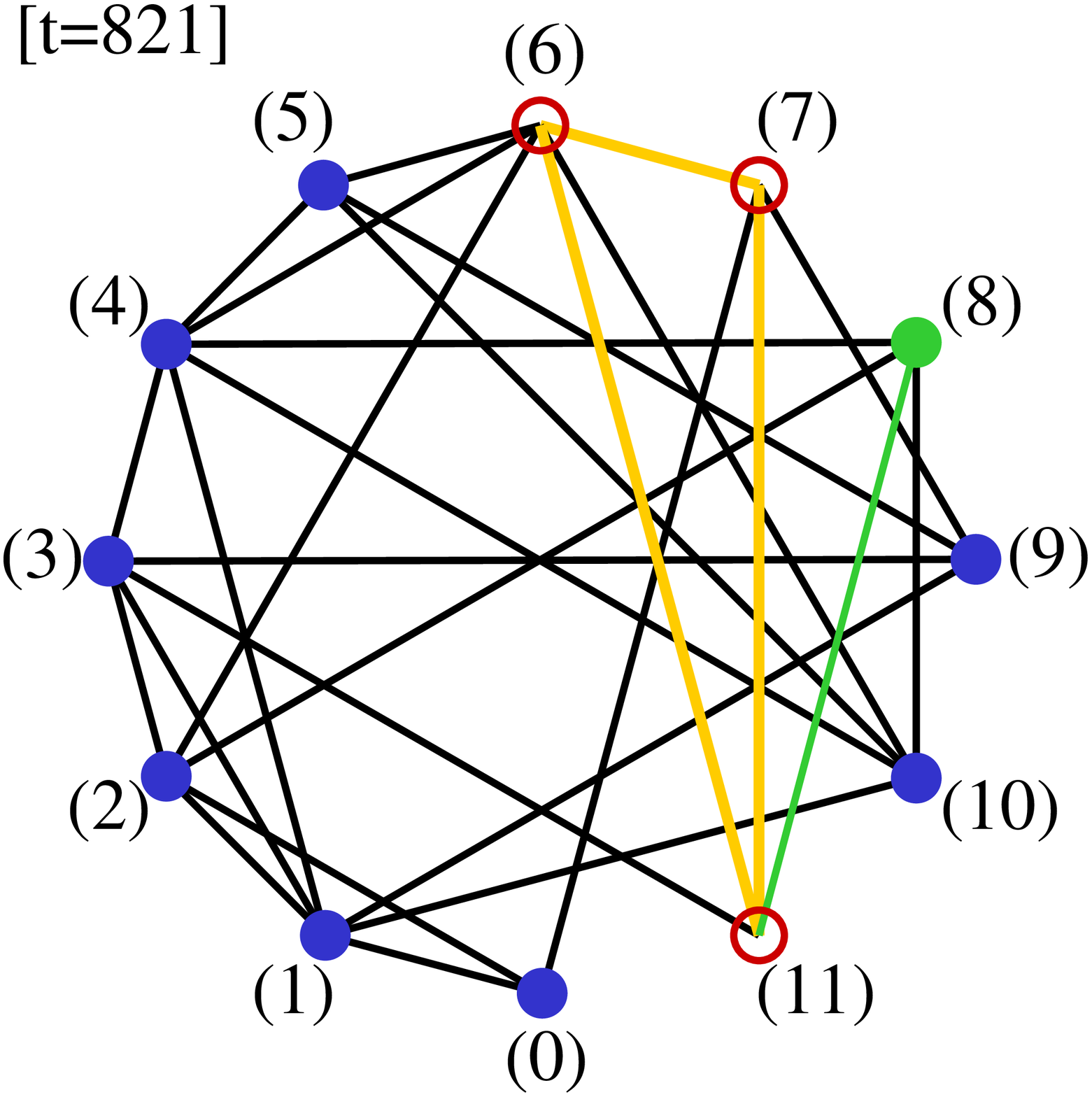,width=0.25\textwidth,angle=0} \hspace{1ex}
\epsfig{file=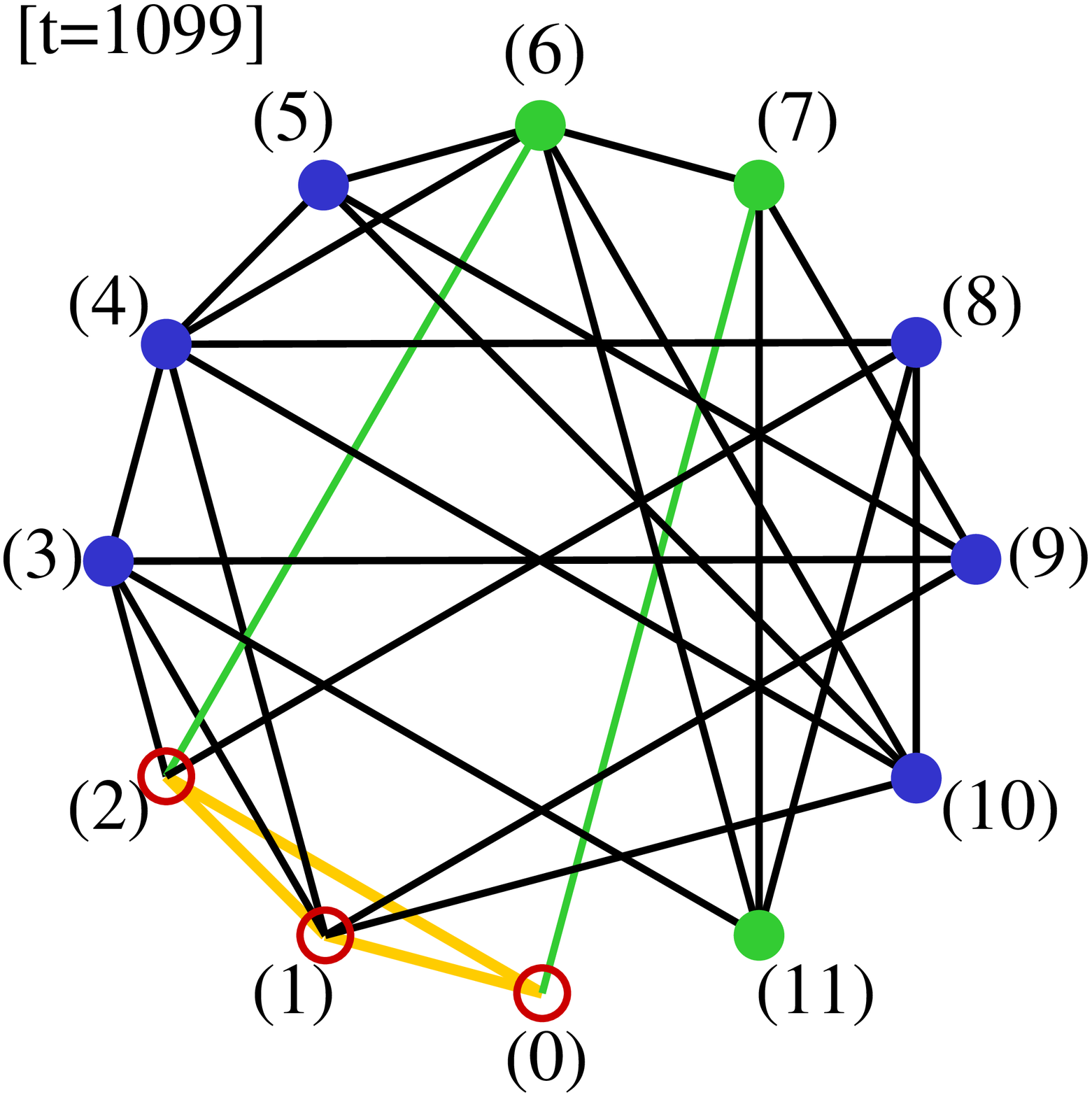,width=0.25\textwidth,angle=0} 
           }
\caption{\small The thought process
$(2,6)\rightarrow(4,5,6,10)
\rightarrow (4,8,10)\rightarrow(8,11)
\rightarrow (6,7,11)\rightarrow(0,1,2)$
of a 12-site network with 7 2-center, 7 3-center and
one 4-center memory state. The non-zero excitatory links 
$w_{i,j}>0$ differ from the uniform level $w$ randomly
by at most 5\%. Compare Fig.\ \ref{fig_12_AA} for
the time-evolution of the variables
		  \label{fig_12_pics}
        }
\end{figure}
%%%%%%%%%%%%%%%%%%%%%%%%%%%%%%%%%%
%%%%%%%%%%%%%%%%%%%%%%%%%%%
%%%%%%%%%%%%%%%%%%%%%%%%%%%

\section{Dynamical Thought Processes}
\label{Dynamical_thought_processes}

In Fig.\ \ref{fig_12_AA} and
Fig.\ \ref{fig_12_pics} we present an
autonomous thought process within a 12-center
network with 15 stable memory states,
illustrated in Fig.\ \ref{fig_illustration}. 
We have chosen a small network here to discuss
the properties of the dynamical thought process
in detail. The model is however completely scalable
and we have performed simulations of networks
containing several thousands of sites without any problem
on standard computers. The choice of synchronous
or asynchronous updating procedures is arbitrary
here, due to the continuous-time formulation. No
particular care needs to be taken for the integration
of the differential equations (\ref{xdot}) and (\ref{phidot})
as the dynamical process has relaxational properties
for the short-time dynamics of both the activities
$x_i(t)$ as well as for the reservoir levels $\varphi_i(t)$.
The model is numerically robust. The dynamics is numerical
stable also for the long-time evolution
of the activities $x_i(t)$ which is driven by the
respective reservoir levels via the reservoir-functions
$f(\varphi)$ and $g(\varphi)$.

The simulations presented in Fig.\ \ref{fig_12_AA} 
were performed using the two distinct
parameter sets listed in Table\ \ref{tab_parameters}.
For the parameter set (a) we observe very rapid
relaxations towards one of the memory-states encoded
in the link-matrices of the network, as shown in
Fig.\ \ref{fig_illustration}. 
The reservoir levels are depleted very slowly
and the memory-stable becomes unstable and 
a different memory-state takes over only
after a substantial time has passed.
Comparing the thought process shown in
Fig.\ \ref{fig_12_AA} with the network of excitatory links 
of the dHAN shown in Fig.\ \ref{fig_illustration} one
can notice that (I) any two subsequent memory states
are connected by one or more links and that (II)
excitatory links have in fact a dual functionality:
To stabilize a transient memory state and to associate
one memory state with a subsequent one. The model
does therefore realize the two postulates 
for self-regulated associative thought processes
set forth in the introduction.

Biologically speaking it is a `waste of time' if individual
memory states remain active for an exceedingly
long interval. The depletion-rate $\Gamma_\varphi^-=0.009$
for the reservoir-levels is very small for the parameter set (a) 
listed in Table\ \ref{tab_parameters}. For the
parameter set (b) we have chosen a substantially
larger depletion-rate $\Gamma_\varphi^-=0.02$
together with very smooth reservoir-functions
$f(\varphi)$ and $g(\varphi)$ and a finite
value for $g^{(min)}$ in order to avoid
random drifts for centers active over prolonged
periods and fully depleted reservoir levels.

%%%%%%%%%%%%%%%%%%%%%%%%%%%%%%%%%%
\begin{figure}[t]
\centerline{
\epsfig{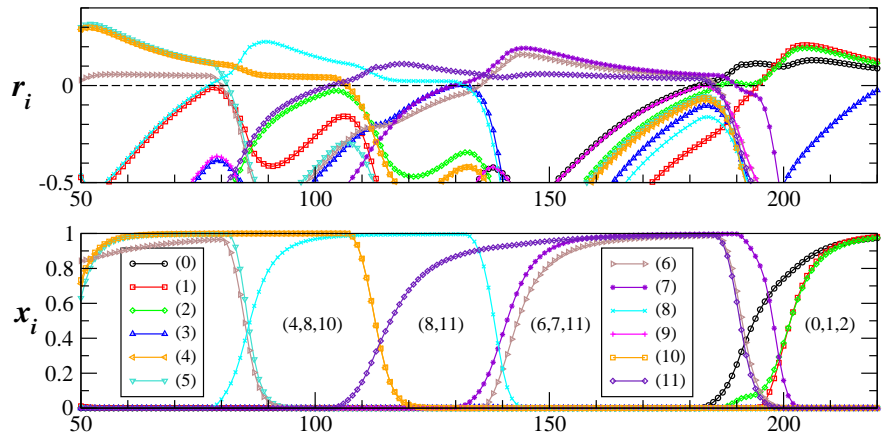} 
           }
\smallskip
\caption{\small The growth-rates $r_i(t)$ (top) and
the activity $x_i(t)$ (bottom) 
for the thought process
$(4,5,6,10)
\rightarrow (4,8,10)\rightarrow(8,11)
\rightarrow (6,7,11)\rightarrow(0,1,2)
$ for the 12-site network shown in Fig.\ \ref{fig_illustration},
using the parameter set (b) of Table\ \ref{tab_parameters}
		  \label{fig_12_RA}
        }
\end{figure}
%%%%%%%%%%%%%%%%%%%%%%%%%%%%%%%%%%

The two thought processes shown in Fig.\ \ref{fig_12_AA}
are for identical link-matrices, only the parameters
entering Eqs.\ (\ref{xdot}) and (\ref{phidot})
differ. We note that the sequence of memory-state
is identical for both parameter sets, the dynamics
is stable over a wide range of parameter-values.
The history of memory states goes through a cycle,
as it is evident for the simulations using the parameter set
(b), since the phase-space is finite. The
12-site cluster used in this simulation contains 15
different memory states and the cycle runs over
6 distinct states, a substantial fraction of the total.
For larger networks with their very high numbers of stored
memory-states 
%(see Buhmann, Divko \& Schulten, 
\cite{buhmann89}
the cycle length will be in general very long for
any practical purposes. We have, however, not yet
performed a systematic study of the cycle-length
on the system properties along the same lines
usually done for random boolean networks
%(Schuster, 
\cite{schuster01}. 

We note that binary cycles do not occur for the 
parameters used here. We have chosen 
$\Gamma_\varphi^+< \Gamma_\varphi^-$ and
the reservoir levels therefore take a while
to fill-up again. Active memory states can therefore
not reactivate their predecessors, which have necessarily
low reservoir levels and small values for $g(\varphi)$.
The same temporal asymmetry could be achieved by
choosing $w_{i,j}\ne w_{j,i}$. We do not rule-out
the use of asymmetric link-matrices for the dHAN, but
it is an important property of the dHAN to be able
to establish a time direction autonomously. 
The types of memory-states storable in the dHAN would
otherwise be limited to asymmetric states.

%%%%%%%%%%%%%%%%%%%%%%%%%%%%%%%%%%
  \begin{figure}[t]
  \centerline{
  \epsfig{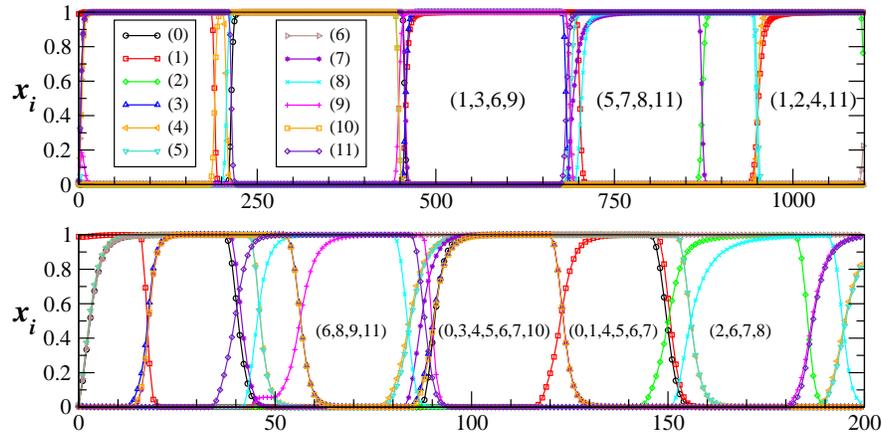} 
             }
  \smallskip
  \caption{\small The activity $x_i(t)$
  for two 12-center clusters with a high density
  of links using the parameter sets (a) and (b) 
  of Table\ \ref{tab_parameters} (top/bottom).
  \newline
  Top: The thought process is
  $(1,4,6,7,11)
  \rightarrow (0,6,7,8,10)\rightarrow(1,3,6,9)
  \rightarrow (5,7,8,11)\rightarrow(2,5,8,11)
  \rightarrow (1,2,4,11)
  $.
  \newline
   Bottom: The thought process is
   $(0,3,4,5,6,7,10)
   \rightarrow (3,6,8,10,11)\rightarrow(6,8,9,11)
   \rightarrow (0,3,4,5,6,7,10)\rightarrow(0,1,4,5,6,7)
   \rightarrow (2,6,7,8)
   $
    \label{fig_12_AA_big}
        }
 \end{figure}
%%%%%%%%%%%%%%%%%%%%%%%%%%%%%%%%%%

%%%%%%%%%%%%%%%%%%%%%%%%%%%
%%%%%%%%%%%%%%%%%%%%%%%%%%%

\section{Details of the Dynamics}
\label{Details_of_the_dynamics}

In Fig.\ \ref{fig_12_RA} we present a blowup
of the thought process presented in Fig.\ \ref{fig_12_AA}
for the parameter set (b), together with the
time-dependence of the growth rates $r_i(t)$.
We can clearly observe how competition among the
ACs plays a crucial role. For the first transition 
occurring at $t\approx 78$ sites (1) and (8) 
compete with each other. Both have two excitatory 
links with the active cluster (4,5,6,10), see
Fig.\ \ref{fig_illustration} and
Fig. \ref{fig_attention}. This competition is
resolved here by small random differences in between
the excitatory links $w_{i,j}$ used here.

The transition occurring at $t\approx190$ 
in between the two disjunct
memory states (6,7,11) and (0,1,2)
takes a substantial amount of time to complete as it
goes through the intermediate state (0,7).
The memory state (0,7) is actually a valid
memory state by itself but is not stabilized here
as the AC (7) looses out in the competition
with (1) and (2), compare Fig.\ \ref{fig_12_RA}.

Note that in the transition
$(4,8,10)\to(8,11)\to(6,7,11)$ this effect does
not occur. The intermediate state (8,11) is
stabilized because the reservoir level of
(6) had yet not been refilled completely due
to a precedent activation. In this case
(6) and (7) loose out in competition with (8).

  In Fig.\ \ref{fig_12_AA_big} we show the results
  of two simulations on different 12-site clusters
  with very dense link-matrices $w_{i,j}$ which
  contain memory-states with up to seven centers.
  Simulations for both set of parameters are shown
  and we observe that the dynamics works perfectly
  fine. With the appropriate choice Eq.\ (\ref{eq_Z})
  for the link-strength even networks with substantially
  larger memory states allow for numerical
  stable simulations.

\begin{figure}[t]
\centerline{
\epsfig{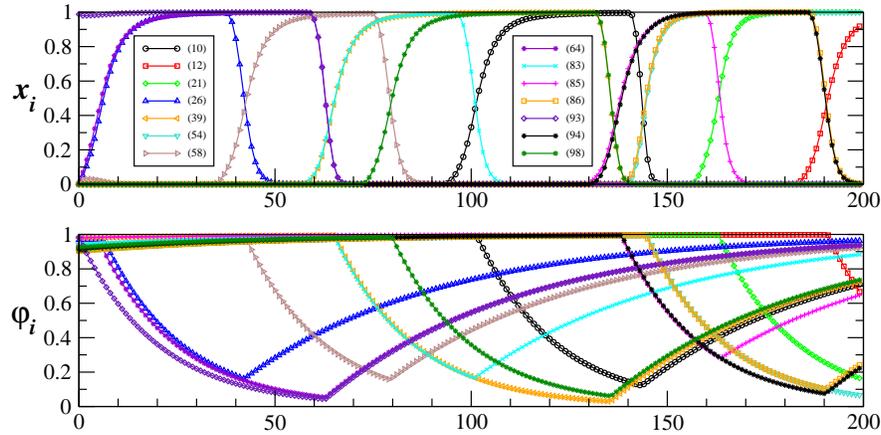} 
           }
\smallskip
\caption{\small The activity $x_i(t)$ and
the reservoir $\varphi_i(t)$ for
a 99-center clusters with 165/143/6 stable
memory states containing 2/3/4 centers,
using the parameter set (b) 
of Table\ \ref{tab_parameters} 
The thought process is
$(26,64,93)
\rightarrow (58,64,93)\rightarrow(39,58,83)
\rightarrow (39,83,98)\rightarrow(10,39,98)
\rightarrow (54,85,86,94) \rightarrow (21,54,86,94)
\rightarrow (12,21,54)
$
		  \label{fig_99_AI}
        }
\end{figure}

Finally we present in Fig.\ \ref{fig_99_AI}
the activities and reservoir-levels
for a 99-site network, using the parameter set
(b). Plotted in Fig.\ \ref{fig_99_AI} are only
the activities and the reservoir-levels
of the sites active during the
interval of observation. We can nicely observe
the depletion of the reservoir-levels for the
active centers and the somewhat slower recovery
once the activity falls again. Similar results
are achieved also by simulations of very big networks. 

%%%%%%%%%%%%%%%%%%%%%%%%%%%%%%%%%%
%%%%%%%%%%%%%%%%%%%%%%%%%%%%%%%%%%

%%%%%%%%%%%%%%%%%%%%%%%%%%%
%%%%%%%%%%%%%%%%%%%%%%%%%%%

\section{Discussion}

We have here investigated the autonomous
dynamics of the dHAN and neglected any
interaction with the outside world.
Sensory inputs would add to the
growth rates by appropriate time-dependent
modulations of the respective bias $b_i(t)$,
see Eq.\ (\ref{ri}). For any sensory input
the bias of the involved activity centers
would acquire a finite positive value during
the interval of the sensory stimulation.

Taking a look at the growth rates plotted
in Fig.\ \ref{fig_12_RA} it becomes immediately
clear that the sensory input needs in general
a certain critical strength in order to influence
the ongoing thought process. The autonomous
thought process and the sensory input compete with
each other. This kind of competition does not
occur in simple attractor neural networks
%(Schuster, 
\cite{schuster01},
for which any non-zero input induces the network to
flow into the nearest accessible attractor. Both for
inputs resembling closely a previously
stored pattern as well as for random and 
nonsensical inputs.

\begin{figure}[t]
\centerline{
\epsfig{file=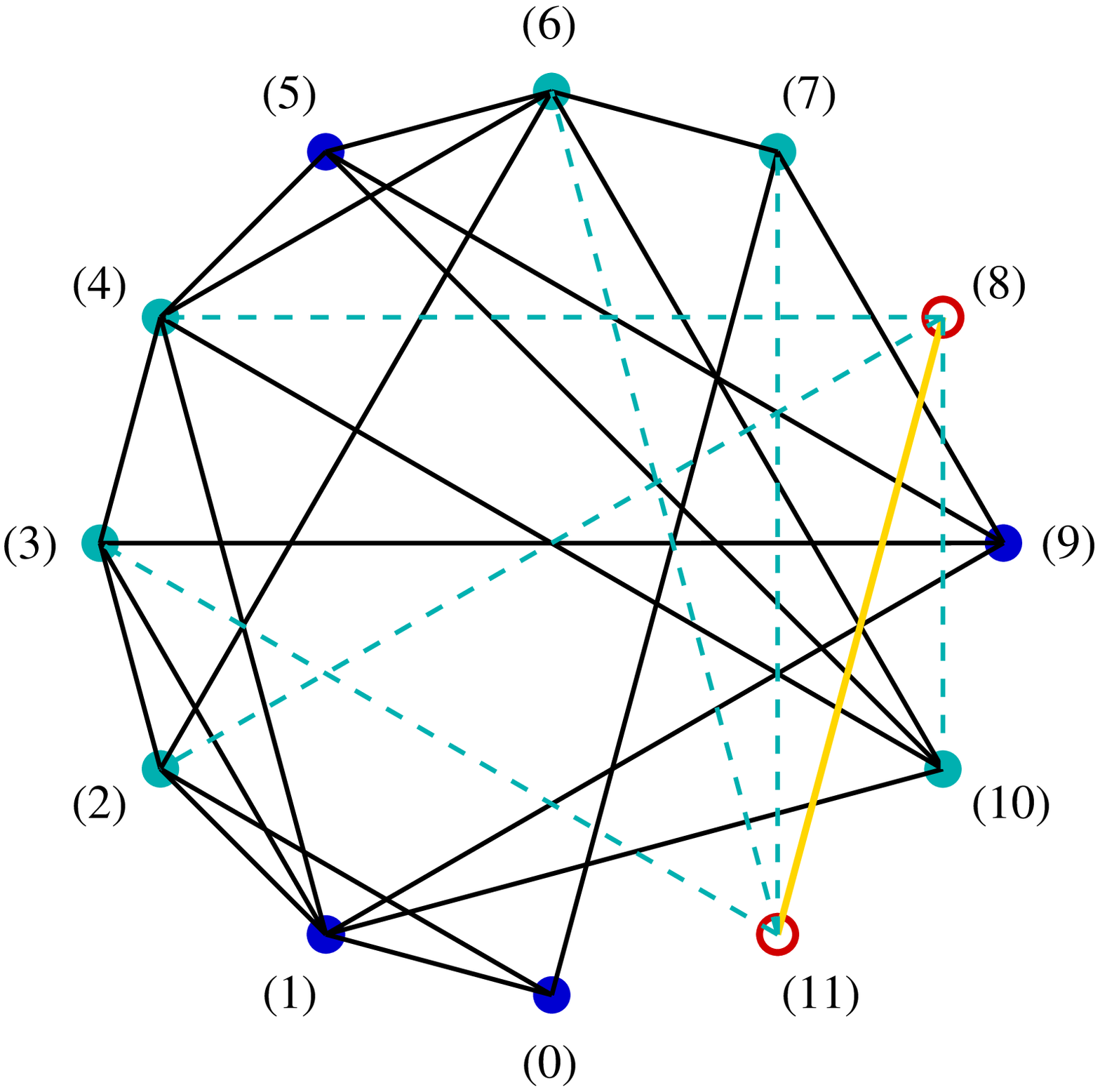,width=0.35\textwidth,angle=0}\hspace{5ex}
\epsfig{file=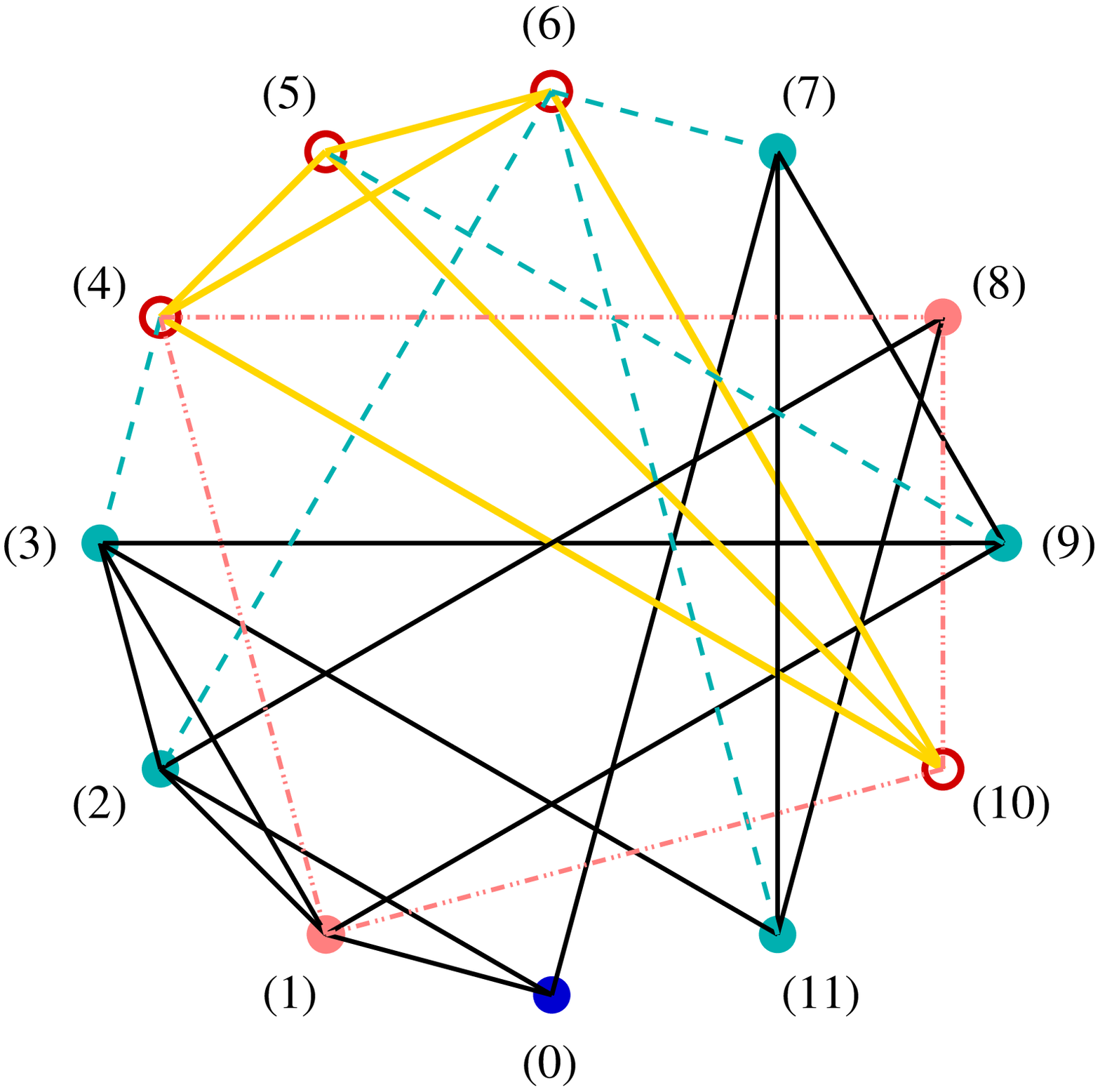,width=0.35\textwidth,angle=0}
           }
\caption{\small Dynamical attention-focusing in
the 12-site network with the thought process
illustrated in Fig.\ \ref{fig_12_RA}, for two
different stable memory states. The dashed/dashed-dotted
cyan and pink
links/circles denote centers linked weakly/strongly 
(by one/two excitatory links) to the active memory state
		  \label{fig_attention}
        }
\end{figure}

The dHAN considered here will, on the other hand,
recognize external patterns only when the input is
such that it wins the competition with the ongoing
thought process, leading to the activation of the
corresponding memory state. The strength of the
sensory input necessary for this recognition process
to be completed successfully depends
crucially on the number of links between the
current active memory state and the ACs 
stimulated by the sensory input, as one
can see in Fig.\ \ref{fig_12_RA}
and Fig.\ \ref{fig_attention}. The sites
(1) and (8) have two links to the memory
state (4,5,6,10), all other centers have either zero
or just one. Any sensory input arriving 
on sites (1) or (8) would be recognized
even for small signal-strength when (4,5,6,10) 
is active, a sensory input arriving at
site (0) would need to be, on the other hand,
very large. This property of the dHAN is
then equivalent to the capability of focus
attention autonomously, an important precondition
for object recognition by cognitive systems
%(see Reynolds \& Desimone, 
\cite{reynold99}.
Every active memory
states carries with it an `association cloud'
and any external input arriving within this area of ACs 
linked directly to the active memory state
will enjoy preferential treatment by the dHAN.

%%%%%%%%%%%%%%%%%%%%%%%%%%%%%%%%%%
%%%%%%%%%%%%%%%%%%%%%%%%%%%%%%%%%%

%%%%%%%%%%%%%%%%%%%%%%%%%%%
%%%%%%%%%%%%%%%%%%%%%%%%%%%

\subsection{Biological Considerations} \label{Biological considerations}

\begin{figure}[t]
\centerline{
\epsfig{file=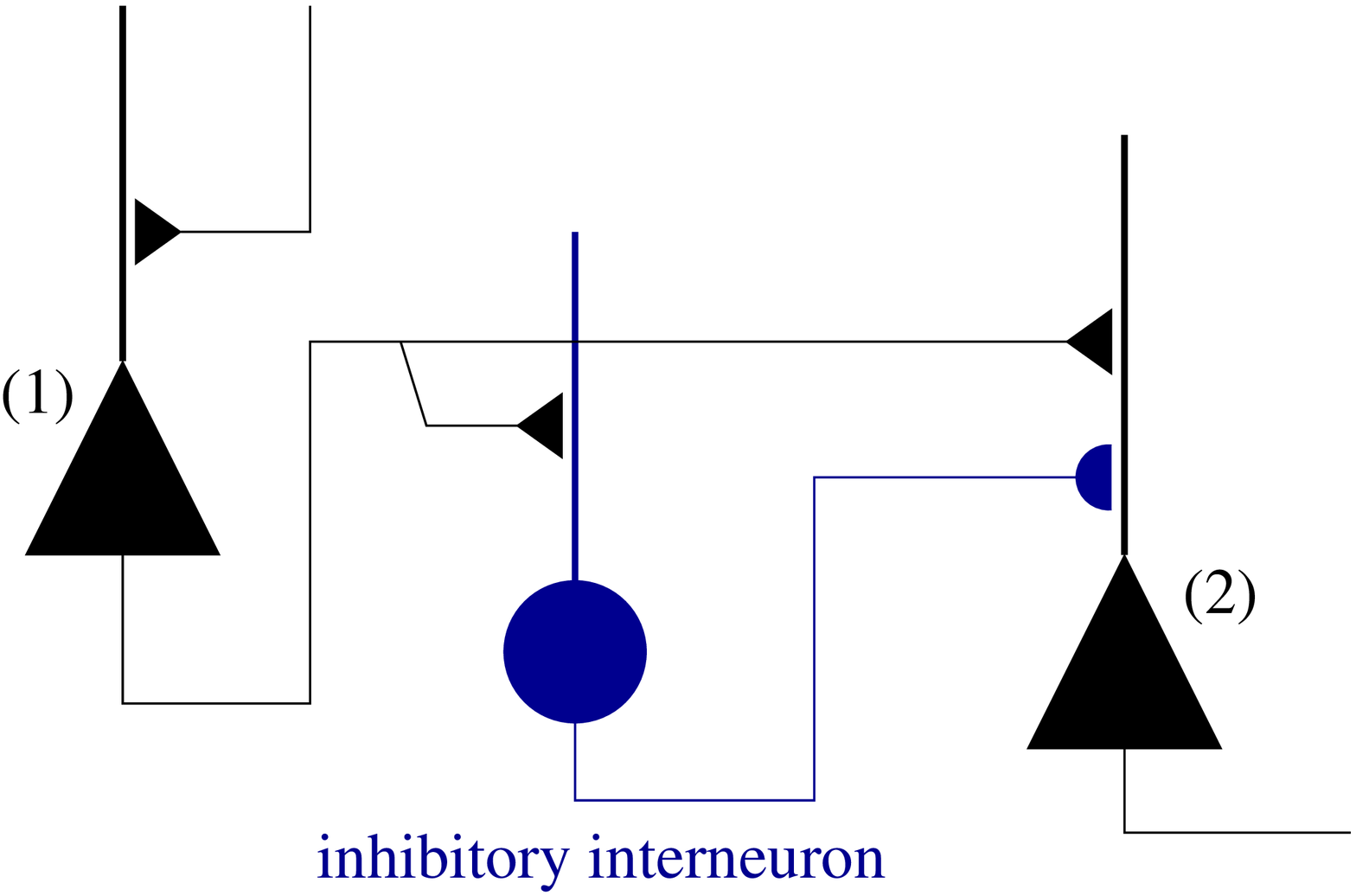,width=0.55\textwidth,angle=0}
\hspace{6ex}
\epsfig{file=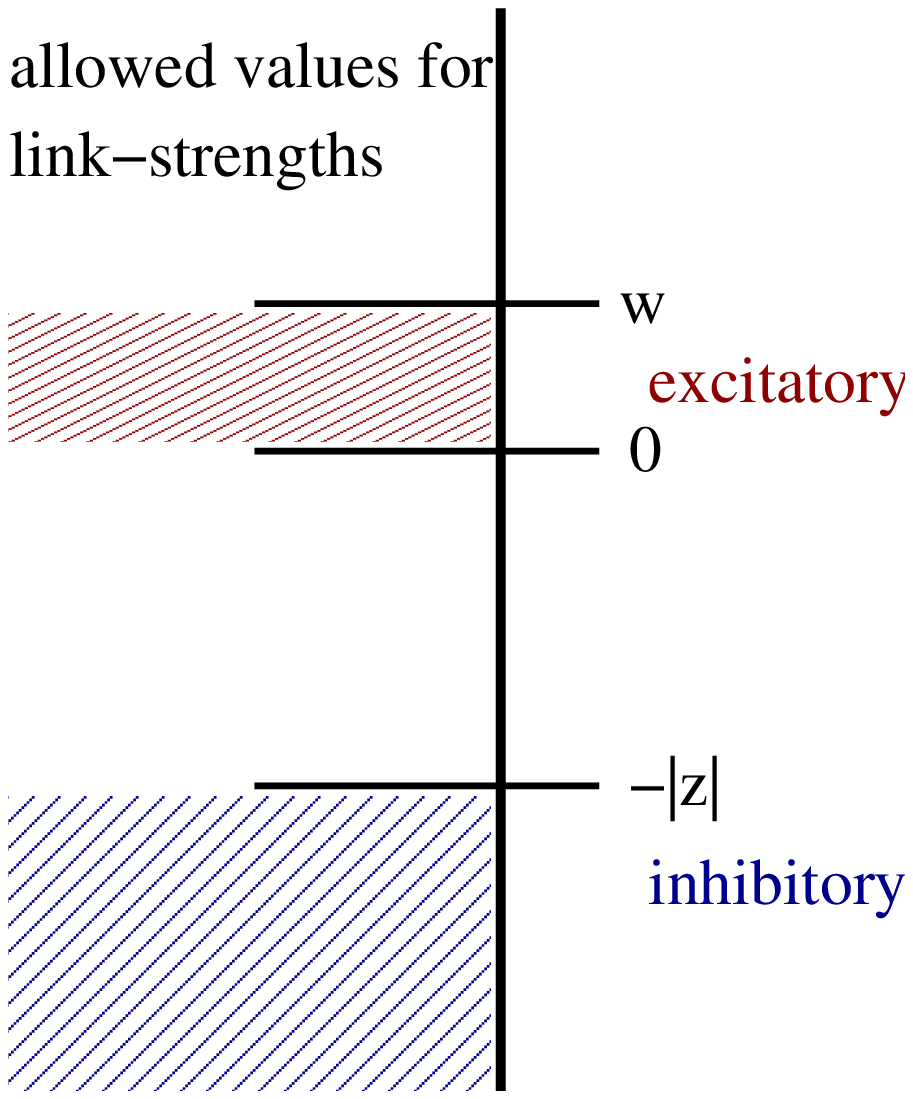,width=0.3\textwidth,angle=0}
           }
\caption{\small Illustration of a network of interacting
neurons leading to
an effective discontinuous inter-neural coupling.\newline
Left:
The excitatory neuron (1) connects both directly to
neuron (2) and indirectly via the inhibitory interneuron.
An activation of the interneuron by neuron-(1) may lead
to complete inhibition of neuron-(2), masking 
completely the direct excitatory (1)-(2) link.\newline
Right: The resulting range for the allowed link-strengths
for the excitatory links $w_{i,j}$ (red shaded region) and for the 
inhibitory links $z_{i,j}$ (blue shaded region)
		  \label{fig_inhibition}
        }
\end{figure}

At first sight the model Eq.\ (\ref{xdot})
possesses an unbiological feature. Neglecting the $f(\varphi)$
and $g(\varphi)$ for a moment, the total effective
link strength $w_{i,j}+z_{i,j}$ is discontinuous:
Either strongly negative ($w_{i,j}=0,\ z_{i,j}\le-|z|$),
or weakly positive ($0\le w_{i,j}\le w,\ z_{i,j}=0$),
as illustrated in Fig.\ (\ref{fig_inhibition}).
This property of the link-matrices between the
constituent activity centers is crucial for the
whole model. It is essential for the stability
of the individual memory states, see
in Sect.\ \ref{Memory_states},
and it forms the 
basis for hierarchical object representations, 
as discussed in Sect.\ \ref{sect_hierarchical_memory_states}.
It constitutes a key difference between our and
other models of neural networks 
%(Schuster, 
\cite{schuster01}. 

The effective link-strength $w_{i,j}+z_{i,j}$ does 
however not correspond to the bare synapsing-strength in
biological neural assemblies. It represents
an effective coupling in between local or
distributed centers of neural activities and 
this kind of discontinuous behavior 
may actually result quite naturally from a simple coupling 
via intermediate inhibitory interneurons, as illustrated 
in Fig.\ (\ref{fig_inhibition}).
When the interneuron is active, the effective coupling 
is strongly inhibitory. When the interneuron is quiet,
the coupling is weakly excitatory with Hebbian-type learning.
When the interneuron is active, it might as well
inhibit the target neuron completely.
Integrating out the intermediate inhibitory interneuron
leads in this way to an effective discontinuous 
inter-neuron coupling.

Biologically speaking, one observes that inhibitory synapses 
are placed all over the target neurons, on the dendrites, 
the soma and on the axon itself. 
Excitatory synapses are, however, located mainly on the
dendrites. This observation suggests that inhibitory synapses
may indeed preempt the postsynaptic excitatory potentials, giving
functionally rise to a model like the one proposed here.

%%%%%%%%%%%%%%%%%%%%%%%%%%%%%%%%%%
%%%%%%%%%%%%%%%%%%%%%%%%%%%%%%%%%%
%%%%%%%%%%%%%%%%%%%%%%%%%%%
%%%%%%%%%%%%%%%%%%%%%%%%%%%

\section{Conclusions}

We have proposed, discussed and implemented
a central principle for the self-generated
time-series of mental states of a cognitive
system - the notion of ``duality'' for memory 
states. A cognitive system can make use of
the stored information - the memories -
only if these memories can be related to
each other. We believe that no separate
conjunction units are necessary for this
job. Memory states and conjunction units 
responsible for the time-evolution of the 
thought process are
- in our view - just two different aspects
of the same coin: Memory states can either be
activated, just as normal memory states are
supposed to behave, or act as associative
links, enabling the dynamics of the thought process.

There may exist a range of possible implementations
of this principle, here we have shown that
an associative network with overlapping memory
states of a generalized neural-network-type
will autonomously generate a history of transient
memory states when suitable couplings to local
reservoir levels are introduced. This 
self-regulating model exhibits, to a certain extent,
autonomous data-processing capabilities with
a spontaneous decision-making ability. 
Here we have been concerned with showing the feasibility
of this approach. Further research will be
necessary to show that this self-regulating
system can carry out specific cognitive tasks.

I thank R. Valent\'\i\ for a careful reading
of the manuscript.

%%%%%%%%%%%%%%%%%%%%%%%%%%%%%%%%%%%%%%%%%%%%%%%%%%%%%%%%%%%%%%%%

%%%%%%%%%%%%%%%%%%%%%%%%%%%%%%%%%%%%%%%%%%%%%%%%%%%%%%%%%%%%%%%%

\end{document}